\documentclass[oldversion]{aa}  
\usepackage{graphicx}
\usepackage{txfonts}
\begin{document}
   \title{A peculiarity of metal-poor stars with planets ?}


   \author{M. Haywood
          \inst{1}
          }

   \offprints{M. Haywood}

   \institute{GEPI, Observatoire de  Paris, CNRS, Universit\'e Paris Diderot; 92190 Meudon,France\\
              \email{Misha.Haywood@obspm.fr}
            }

   \date{Received September 15, 1996; accepted March 16, 1997}

 
  \abstract  
{Stars  with   planets  at   intermediate  metallicities
([-0.7,-0.2]  dex) exhibit  properties  that differ  from the  general
field stars.  Thirteen stars with planets reported in this metallicity
range  belong to  the  thick disc,  while  only one  planet have  been
detected among  stars of the  thin disc.  Although this  statistics is
weak,  it contradicts the  known correlation  between the  presence of
planet  and  metallicity.  We  relate  this  finding  to the  specific
property of the  thin disc in this metallicity  range, where stars are
shown  to rotate around the Galaxy faster than  the  Sun.  Their
orbital parameters are conveniently explained if they are contaminants
coming from  the outer  Galactic disc, as  a result of  radial mixing.
This must be  considered together with the fact  that metal-rich stars
([Fe/H]$>$+0.1 dex)  found in the  solar neighbourhood, which  are the
hosts  of  most  of  the  detected planets,  are  suspected  of  being
wanderers from  the inner Galactic  disc.  It is then  questionned why
stars  that  originate   in  the  inner  and  outer   thin  disc  show
respectively the highest  and lowest rate of detected  planets.  It is
suggested  that the  presence of  giant planets  might be  primarily a
function  of a  parameter  linked to  galactocentric  radius, but  not
metallicity. Combined  with the existing  radial metallicity gradient,
then  radial  mixing  explains  the correlation  at  high  metallicity
observed locally,  but also the peculiarity found  at low metallicity,
which  cannot  be  accounted  for  by  a  simple  correlation  between
metallicity  and  planet  probability.  
}  


   \maketitle
%

\section{Introduction}

The parent  stars of  exoplanets are  known to be,  on the average, more
metal-rich than  'single' stars  of the solar  neighbourhood (Gonzales 1997;
Santos, Israelian \& Mayor 2000).  
Apart from this peculiarity, stars with planets show little,
if  any,  specific  properties (see Udry \& Santos (2007) for a review).   They  do not seem  to  have  peculiar
chemical  abundance ratios  (Ecuvillon et  al., 2006),  although slight
differences  have  been  claimed   (Robinson  et  al., 2006),  while
kinematic  behaviour  also  seems  similar  to that  of  field  stars
(Barbieri  \& Gratton, 2002).   The origin  of  this preference  for
metal-rich stars is still discussed,  with two possible hypothesis: primordial
(Santos, Israelian \& Mayor, 2000) or  gained by  the star  because of  a possible
infall episode at its birth (Gonzales 1997; Laughlin 2000). Recent advances, observational
as well as theoretical (Ida \& Lin (2005), Mordasini et al. 2007) seems to some support the former hypothesis.

Radial mixing, combined with the metallicity gradient in the Galactic disc, 
is probably responsible for  the extended
tails  of the metallicity  distribution of  both metal-rich  {\it and}
metal-poor stars of the thin  disc in the solar neighbourhood (Haywood 2007).
Given that a  majority of planets are  detected around
metal-rich stars, which seems to follow the same kinematics behaviour as the
parent metal-rich population, it  implies that most would originate from
the   inner  Galactic  disc (Ecuvillon et al. 2007).    The  intermediate   metallicity  range
([-0.7,-0.2]dex) shows a more  complex pattern because it involves two
distinct  populations:  the  thick   disc  at
[Fe/H]$<$-0.2, and the thin disc starting at about
[Fe/H]=-0.7  upward. However, these  two populations
differ on several points. First, they  have distinct kinematic
behaviour, the  thick disc lagging  the Local Standard of Rest (LSR), while the  thin disc metal-poor
population have V component of space motion distinctly higher than average (see next section). Second, 
they have  different level of $\alpha$ enrichment,  with the thick
disc at [$\alpha$/Fe]$>$0.17 dex and  the thin disc between 0. and 0.1
dex. Finally, they  probably have distinct age distributions. The thick
disc  or transition stars  are essentially  old objects (ages $>$8 Gyr),
while thin disc  metal-poor  stars as  young  as  2 Gyr  can  be  found, even though
most are older than this lower limit.  These
properties, and the hiatus  in metallicity between the two populations
(the fact  that an  old population,  the thick disc,  has stars  up to
[Fe/H]=-0.2, while  a younger  one, the thin  disc, has stars  down to
[Fe/H]$\approx$-0.7),  have  been interpreted  in  Haywood (2007)  has
testifying  the  different origins  of  these  two  populations :  the
metal-poor thin disc stars being wanderers  from the outer  Galactic disc,
while  the  thick  disc  and  transition stars  stem  from  the  
evolution of the disc at the solar circle.

In the present study, we examine orbital and chemical characteristics of planet 
host stars to see how they follow this general picture. In the next section, we first
show evidences of the radial mixing effects on solar neighbourhood stars, then 
compare the chemical and kinematics parameters of stars with planets with the general population.
We discuss our results in the last section.


\section{A peculiarity of metal-poor stars with planets}

\begin{figure}
\includegraphics{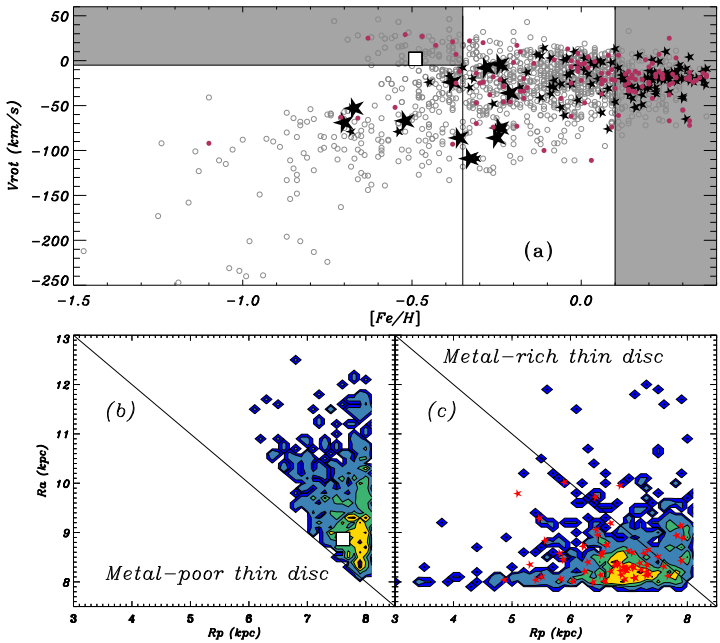}
\caption{{\bf (a)} ([Fe/H], V$_{rot}$) distribution for stars 
in the samples of Reddy et al. (2003), Reddy et al. (2006), Bensby et al. (2005) and Valenti \& Fischer (2005).
Metal-poor stars (-0.8$<$[Fe/H]$<$-0.3 dex) in this plot separate in two groups : one made of thick disc stars lagging 
the sun with V$_{rot}$ within [-40,-80] km/s.  The other one is
preceding the Sun and extending towards  ([Fe/H]=-0.8, Vrot=+60km/s).
Stars with planets from the sample of Gilli et al. (2006) are shown as star symbols, the larger symbols
representing the objects of Tab. 1. Stars with no detected planet from Gilli et al. (2006) are represented 
by dots. The only star detected with a planet in the metal-poor thin disc group (HD 171028) is represented by a square in panel (a)
and (b).
The grey areas give the limits of the two subsamples of thin disc stars with (1) [Fe/H]$<$-0.35 and  V$_{rot}>$-5 km/s.
and (2) stars with [Fe/H]$>$0.1 dex. 
Panels {\bf (b)} and {\bf (c)} show the distribution of apo- and peri-centers ($R_{a}$, $R_{p}$) 
for stars in the GCS catalogue within the same limits. 
Stars with planets and [Fe/H]$>$0.1 dex  are shown as star symbols in panel (c). 
The line on plot (b) and (c) 
materialises an orbit with mean ($R_{a}$+$R_{p}$)/2=8kpc.
}
\label{rmeffect}
\end{figure}

\subsection{Evidences of radial mixing from solar neighbourhood stars}

Thin disc  stars at the two  extremes of the  metallicity interval have
markedly different kinematic behaviour. This is easily seen in a plot
of the V velocity component (with respect to the Sun) as a  function of  metallicity
(Fig. 1).  Fig. 1a shows a  distribution of these  two parameters from
various samples  with accurate measurements  of metallicity :  Reddy et
al.  (2003), Reddy, Lambert \& Allende Prieto (2006),  Bensby  et   al.  
(2005),  Valenti   \&  Fischer
(2005).  These  data sets  have  been  cross-identified  with  the  Geneva-Copenhagen Survey (hereafter GCS)
(Nordstr\"om  et  al.   2004)  in  order  to  obtain  the
V velocity component of each  star.  Fig. 1a shows  that the
metal-poor population (which we  define here conservatively as [Fe/H]$<$-0.35 dex) of
the disc  separates in two groups.
The first one is mostly preceding the
LSR, and  limited in  metallicity to -0.8$<$[Fe/H]$<$-0.35 dex (grey area on Fig. 1a),
with probably few stars only having higher metal content.  
 In the GCS catalogue, stars rotating
with  the  LSR have  a  mean metallicity  of  -0.16  dex, while  those
preceding the LSR  by 5 km.s$^{-1}$ have a  mean metallicity of -0.24
dex,  -0.31 dex at  10 km.s$^{-1}$,  and -0.38 dex  at 20  km.s$^{-1}$. 
The second group is lagging the  LSR, the lag
increasing with  decreasing metallicity. 
These two groups
are easily  identified as the thin  disc and the thick disc, and also
correspond  to  stars with  distinctive  $\alpha$-element ratios  (section 2.2.2). 

Because of the position of the Sun at about 8kpc from the galactic center, 
its vicinity is likely to be contaminated by both metal-rich stars from the inner 
disc and metal-poor stars from the outer disc. 
In order to contrast most clearly the orbital properties of these two groups, 
we now  select them from the GCS  catalogue by imposing V$>$-5km.s$^{-1}$  and [Fe/H]$<$-0.35  dex and
metal-rich stars by  [Fe/H]$>$+0.1 dex (grey areas on Fig. 1a). 
Their apo and  peri-centre distributions  ($R_{a}$, $R_{p}$) are shown on panel (1b) and (1c).
The metal-poor group is confined to the upper part of Fig. 1b due to the 
kinematic criterion, while the metal-rich stars occupy an almost distinct
area. The group  of metal-poor
thin disc stars  populates mainly outer orbits, or  orbits with larger
angular momentum, while metal-rich objects symmetrically occupy orbits
of lower  angular momentum.   The fact that  low and  high metallicity
stars  are so prevalent  among each  of these  two distinct  groups, a
property that is expected from the existence of the radial metallicity
gradient, clearly indicates that they most probably are the signatures
of the radial  mixing processes in the Galactic  disc (see also Famaey
et al. (2007)  about the Hyades stream, which  conforms well with this
general  picture).  In the  scenario presented  by Sellwood  \& Binney
(2002) for example, stars  are supposedly loosing  or gaining angular  momentum at
corotation because of spiral waves.  This is a secular process, and it
takes several gigayears  before stars in the outer  and inner Galactic
disc  reach the solar  orbit.  It  means  that we  may expect  to see  the
signature  of this  process on  stars approaching  the solar  orbit from
the outer or inner disc.  In this case they  must still
appear with orbital  characteristics  slightly off   those of   the  main
population.  Because of the radial metallicity gradient, metal-rich
stars are expected to come  from the inner disc, or to approach the solar circle 
from smaller radii (Fig. 1c),  while metal-poor thin
disc stars would come mainly from the outer thin disc (Fig. 1b).
If the distribution (R$_p$,R$_a$ ) of metal-rich stars with planet 
follows the behaviour of field stars in panel Fig.1c, it can be deduced
similarly that they originate from the inner disc.
Stars with [Fe/H]$>$ 0.1 dex from the sample of Gilli et al. (2006) are
shown as star symbols on Fig. 1c. They are superimposed 
with no noticeable differences with field stars. 
This result must be compared with the distribution of metal-poor thin disc stars of 
panel (b), where only one star with planet has been found (HD 171028).
Since these two groups of stars probably originate from 
very different galactocentric distances, this may be a hint
that a parameter linked to the  distance to the galactic center could play a role
in the presence of planets. We now discuss if metallicity is the right parameter.

\subsection{The paucity of metal-poor thin disc star with planets}

This work is based on the two samples (stars with and without planets)
of Gilli et al. (2006), which is the only study providing measurements of
$\alpha$-element abundances for a significant number of stars of both
categories. It has been completed for a few low-metallicity stars with planets
from the literature and not available in the selection of Gilli et al. (2006).

\subsubsection{Kinematics}

The sample of stars with known planet (star symbols) and stars with no detected planet 
(dot symbols) from Gilli et al. (2006) are shown on the ([Fe/H], V$_{rot}$) distribution 
of Figure \ref{rmeffect}.
The kinematic parameters  are from the GCS catalogue.
We emphasize that all stars known to harbour planets at [Fe/H]$<$-0.20 dex and with available estimates
of the [$\alpha$/Fe] ratio (therefore excluding M dwarfs) are included in the sample, and are shown as large star symbols
on Fig. 1a.
A planet orbiting the metal-poor star HD 171028 has recently been reported (Santos et al. 2007). 
HD 171028 is not an Hipparcos star, and its distance
is uncertain (Nordstr{\"o}m et al. report 90pc, while Santos et al. give 43pc), and so are
its velocities. However, the V component (2km.s$^{-1}$ in the GCS catalogue) combined with its 
metallicity  ([Fe/H]=-0.49 dex) and $\alpha$ element content ([$\alpha$/Fe]=0.05 dex),
confirms that HD 171028 is a metal-poor thin disc star. 
Although the overall statistics is small at [Fe/H]$<$-0.20 dex, it illustrates that stars with planets 
are found essentially in the group lagging the LSR, either in the 'thick disc' population or
intermediate between the thin disc and the thick disc, with only one object in the group of 
metal-poor thin disc stars.

\subsubsection{$\alpha$-element ratio}

Stars of the thin and thick discs show distinct properties 
in their level of $\alpha$ enrichment as a function of metallicity. 
This is illustrated in studies of Fuhrmann (1998) and 
Reddy et al. (2003, 2006) (among others), which samples are displayed in Fig. \ref{fehvrot}.
The hiatus between the metal-poorest stars of the thin disc and the
thick disc is apparent in these two plots (metal-poor thin disc stars at ([Fe/H], $\alpha$/Fe])=(-0.6,0.1) dex
metal-rich thick disc at (-0.2,0.15) dex).
 It is however well explained (Haywood, 2007) because metal-poor thin disc stars (located
on Fig. 2a at [Fe/H]$<$-0.3 dex and [$\alpha$/Fe]$<$0.1 dex) are identified as the stars 
with V$_{rot}>$-5km.s$^{-1}$ in Fig. 1a. In other words, these objects have probably formed outside
the solar circle, and probably are not relevant tracers of the chemical evolution at the solar radius.
If we discarded these stars on Fig 2a, the other objects form a continuous sequence
from the thick disc to the thin disc,  materialised by the continuous curve. 

The  two samples  of stars  with and  without planets (limited to [Fe/H]$<$-0.2  dex) are  plotted on
Fig. 2(a\&b). The dichotomy between the two
is  apparent. To the notable exception of HD 171028, stars with  planets have a tendancy to avoid the  area of
metal-poor thin  disc stars, and are positioned  on the intermediate
path between the thin and thick discs (between -0.3 and -0.2 dex), then
in    the   thick    disc   regime    (at   [Fe/H]$<$-0.3    dex   and
[$\alpha$/Fe]$>$0.15  dex or  [Mg/Fe]$>$0.2 dex).  Note that among stars 
with planets which have the lowest amount of $\alpha$-elements, 
HIP 98714  is
unambiguously classified as a  transition or old thin disc stars (not
as a  metal-poor thin disc with  outer-disc origin),  according to its
lag on the V component (-48 km/s), the same for HIP 64459 (at -86 km/s), while
HIP     3479     is     more    intermediate     ([Fe/H]=-0.24     and
V$_{lag}$=-4km.s$^{-1}$) between the two groups.

\subsubsection{The age of metal-poor stars with planets}

\begin{figure}
\includegraphics{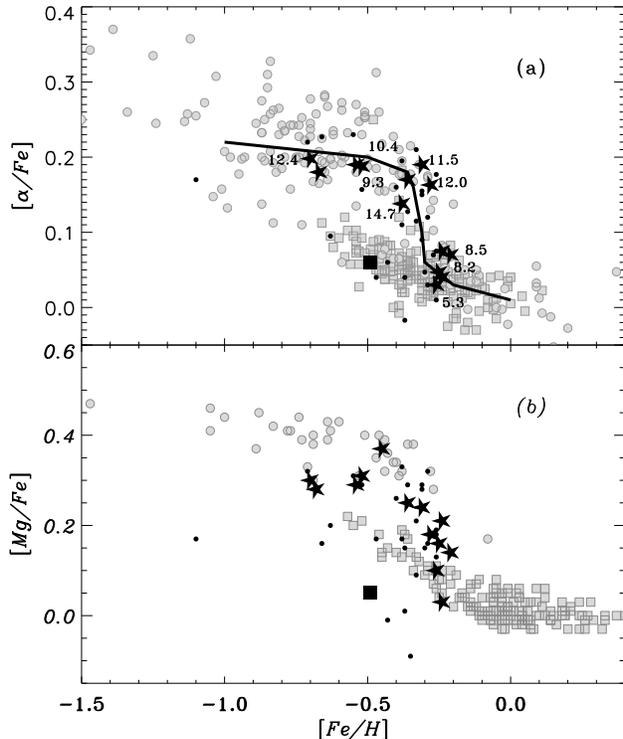}
\caption{[$\alpha$/Fe] (a) and [Mg/Fe] (b) as a function of [Fe/H] for stars with planets considered in this paper (star symbols).
Stars with no detected planet are represented by dots. The square symbol represents the position of HD 171028. 
Grey symbols (squares or circles for kinematically defined thin or thick discs respectively as defined by each study) are field stars from
Reddy et al. (2003, 2006) on plot (a), and Fuhrmann (1998, 1999, 2004) on plot (b). The thick line is only indicative 
and illustrates the evolutionary sequence from the thick disc to the thin disc.
}
\label{fehvrot}
\end{figure}

Stars with planets in the metallicity range of interest here and which
could  be  dated  are   shown  on  Fig. \ref{fehvrot}a  with  their
ages (see also Tab. 1). Stars  were dated using the procedure  described by J{\o}rgensen
\& Lindegren (2005) with the  isochrone set of Demarque et al. (2004), atmospheric parameters from Tab. 1
and absolute magnitudes from Hipparcos parallaxes.
HD 171028 could not be dated to useful precision (due to the uncertainty of its parallax).
The  age
distribution  of stars  with  planets does not  contradict the  general
scheme just presented.  If we adopt the general scenario described
in studies of  solar neighbourhood stars (Bernkopf, Fidler \&
Fuhrmann 2001, Haywood  2006, 2007)
the thin disc has emerged from a transition phase with the thick disc
at an  age of about 8 Gyr.  It is unknown how  long this transition
phase has  lasted, and  ages of thick  disc stars are  uncertain (9-13
Gyr). However, Fig. 2a illustrates that stars with planets roughly follow this sequence, 
from stars of the old thin disc at [Fe/H]$\approx$ -0.2 dex and ages of about 8 Gyr to 
stars in the thick disc regime, enriched in $\alpha$-elements and older by a few Gyr.

\section{Discussion}

We find  there is only 1 star  with planets in the  metal-poor thin disc
population, namely those stars  more metal-poor
than approximately -0.2 to -0.3 dex, having  low level  of $\alpha$-element  enrichment and
mostly preceding the LSR. Note that the approximate  metallicity interval [-0.3,-0.2] dex  may
contain a mixture  of interlopers from the outer  disc and genuine solar
galactocentric stars. Are the above results significant ?
The total number of stars in the two groups (metal-poor thin disc 
and thick disc or intermediate stars, as separated by a limiting 
lag of -5 km.s$^{-1}$) in V, at a metallicity smaller than -0.20 dex\footnote{The limiting metallicity 
adopted in section 2.1 was -0.35 dex. We now move to -0.20 dex because $\alpha$-element
content allow a correct separation of the two groups of stars up to this metallicity.}, are
respectively 9 and 29 stars, with 1 and 13 stars in each group harbouring planets. 
Note that the number of stars with no detected planet in the group of metal-poor thin disc
objects cannot be taken as indicative of the number of surveyed stars in this group, because
no exhaustive list has been published up to now. The consequence is that the 
significance of our result must be evaluated on the number of planet found in each group.

There are, in principle, two different ways to proceed.
If we knew the total number of stars surveyed in each of the two groups (therefore,
the total number of stars with no detected planet), we could estimate whether the 
difference of detected planets (13 and 1) is significant. 
However, these figures are not publicly available. 
Moreover, we would need information on the $\alpha$-element content of the stars, 
(since it is the best criterion to discriminate between the two groups),
but this is measured only on a restricted number of objects.

Another way to proceed is to assume that no selection criterion 
that could favour the inclusion of metal-poor thin disc or thick disc stars
have been adopted in exoplanet surveys.
Then we may estimate the expected relative number 
of stars in each of the two groups, and consecutively infer that, 
if the probability of finding an exoplanet depends only (or mainly)
on metallicity, the relative proportion of detected exoplanet 
in the two groups should follow the same ratio.
The relative normalisation of the two groups of 
stars can be obtained from a distance-limited sample.
The data set of Fuhrmann (2004) is designed to be complete for FGK dwarfs 
up to 25pc. It contains by now about 250 stars (out of approximately 300 objects 
when completed) 
and according to the author it is representative of the final sample. 
8 stars in the sample of  Fuhrmann (2004) clearly belong to the thick disc and have corresponding
level of [Mg/Fe] ($>$0.3 dex) and one is intermediate. 12 stars belong 
to the thin disc metal-poor tail and have [Fe/H]$<$-0.3 dex (which, given the poissonian 
uncertainties, can be considered as equivalent). We assume conservatively an equal
probability of finding a star in the metal-poor thin disc or in the thick disc.
It implies that we should expect the same number of detected planets 
in the two groups.   
What is then the probability that, if the proportion of stars with planets is about the same
in the two groups, 13 out of 14 found so far in this
metallicity range belong to one group only ? The binomial
probability function gives a probability of 8.10$^{-4}$
that this could have occured by chance. 
In other words, at a given metallicity, stars are less likely 
to harbour a giant planet if they originate from the outer Galactic disc than a 
star born at solar galactocentric radius. 
This result must be considered together with the fact that
metal-rich  stars, which are the hosts of most of the detected planets,
have kinematic properties suggesting an  origin
in the inner disc. In other words, stars originating in the outer disc have the lowest 
rate of detected planets, while stars coming from the inner disc show the highest
rate, stars at the solar circle being intermediate between these two extremes.

These  facts   suggest  that, being apparently correlated to metallicity,
the presence  of  giant planets  might be primarily dependent of some specific 
property of the ISM at the original galactocentric distance of the star.
Although theoretical modeling suggests that metallicity is a key parameter (cf. references in the Introduction), 
it is possibly not the single parameter, and perhaps not even a parameter related to the presence of  a planet,
since otherwise we would expect similar proportions 
in the two groups studied here. On the other hand, a parameter more strongly correlated 
with  galactocentric distance than metallicity  must  play a role.   
Giving a possible candidate would be hasardous at this stage, but inspecting the radial variation of 
some specific ingredients (dust ?) of the ISM would be worthwhile.

It must be noted that the apparent correlation 
between metallicity and the detection of planets is a natural consequence of 
these results.
Since the mean metallicity of field stars increases towards the galactic center, planets 
forming preferentially on stars in the inner disc will form 
on metal-rich stars. Radial mixing is then responsible for bringing these stars to the solar circle.
The dependency with some other specific property related to the radial distance to the 
galactic center (other than metallicity) could also explain the peculiarity found at lower metallicity, which cannot be accounted 
for by the simple dependence with metallicity.
If these inferences are correct,  we can also evaluate the proportion of stars
with exo-planets that are truly originate from the solar cicle.
The mean  metallicity of  the  Galactic disc  in  the solar  neighbourhood
is about [Fe/H]=0.0 dex (Haywood 2001), while the terminal metallicity of stars
at solar galactocentric radius  (the upper most metallicity reached by
stars born at  solar radius), is about  0.15 dex (Haywood 2006).
Only half of  the stars  with detected 
planets have a metallicity below this limit, and can be considered as
truly   endemic  to   the  solar circle.

\begin{table}
  \caption{Stars with planets considered in this study.
}
  \begin{center}
    \leavevmode
    \footnotesize
        \small
    \begin{tabular}[h]{llllll}
      \hline  \hline
      Id  &   [Fe/H]    &   T$_{eff}$  &  [$\alpha$/Fe]   & Age & V \\  \hline
      HIP  &    dex    &   K      &  dex   &  Gyr   & km$^{-1}$ \\ \hline
 3479      &   -0.24     & 5626   & 0.04  &  & -4\\
 3497      &   -0.33     & 5636   & 0.19  & 11.5 (6.5, 16.1)   & -109\\
 5054      &   -0.52     & 5835   & 0.188  &  10.4 (5.6, 14.8) & -67\\ 
10138      &   -0.24     & 5163   &  0.075 &  &-75\\
26381      &   -0.38     & 5546   & 0.138  & 14.7 (8.2, 20.0) & -24\\ 
31688      &   -0.54     & 4554   & 0.19  &  9.3 (7.4, 11.2)\\
58952      &   -0.28     & 4773   & 0.163  &  & -8\\
62534      &   -0.36     & 5494   & 0.17  & 12.0 (4.3 19.5)  & -86\\
64426      &   -0.70     & 5884   & 0.198  & 12.4 (9.5, 15.0) & -69\\ 
64459      &   -0.25     & 5886   & 0.047  &  8.2 (6.3, 10.0) & -86\\ 
78459      &   -0.21     & 5853   & 0.07  &  8.5 (5.8, 10.7) & -36\\ 
83949      &   -0.67     & 5868   & 0.18  &  & -53\\  
98714      &   -0.26     & 5327   & 0.03  &  5.3 (4.5, 6.3)  & -48\\ \hline
HD 171028     &   -0.49     & 5663     & 0.05  &                  &  2  \\ \hline
      \end{tabular}
  \end{center}
Notes: Atmospheric parameters from Gilli et al. (2006) or Soubiran \& Girard (2005) for HIP 3497 and HIP 83949.
V velocities from Nordstr\"om et al. (2004). 
Ages are derived as described in the text,
except for HIP 31688, which age comes from da Silva et al. (2006).
The ages of HIP 58952 (giant), HIP 3479 and HIP 10138 could not be determined.
Parameters of HD 171028 from Santos et al. (2007) are given.
\label{tabSTCol}
\end{table}

\begin{acknowledgements}
I would like to thanks Fr\'ed\'eric Arenou, ALM, and the referee for their help and useful comments which much improved
this article.
\end{acknowledgements}


\begin{thebibliography}{}
\bibitem[\protect\citeauthoryear{Barbieri \& Gratton}{2002}]{2002A&A...384..879B} Barbieri M., Gratton R.~G., 2002, A\&A, 384, 879 

\bibitem[\protect\citeauthoryear{Bensby et al.}{2005}]{2005A&A...433..185B} 
Bensby T., Feltzing S., Lundstr{\"o}m I., Ilyin I., 2005, A\&A, 433, 185 

\bibitem[\protect\citeauthoryear{Bernkopf, Fidler, \& 
Fuhrmann}{2001}]{2001ASPC..245..207B} Bernkopf J., Fidler A., Fuhrmann K., 
2001, ASPC, 245, 207 




\bibitem[\protect\citeauthoryear{da Silva et 
al.}{2006}]{2006A&A...458..609D} da Silva L., et al., 2006, A\&A, 458, 609 

\bibitem[\protect\citeauthoryear{Demarque et 
al.}{2004}]{2004ApJS..155..667D} Demarque P., Woo J.-H., Kim Y.-C., Yi 
S.~K., 2004, ApJS, 155, 667 

\bibitem[Ecuvillon et al.(2007)]{2007A&A...461..171E} Ecuvillon, A., 
Israelian, G., Pont, F., Santos, N., Mayor, M.\ 2007, \aap, 461, 171 


\bibitem[\protect\citeauthoryear{Famaey et al.}{2007}]{2007A&A...461..957F} 
Famaey B., Pont F., Luri X., et al., 2007, A\&A, 461, 957 


\bibitem[\protect\citeauthoryear{Fuhrmann}{1998}]{1998A&A...338..161F} 
Fuhrmann K., 1998, A\&A, 338, 161 

\bibitem[Fuhrmann(1999)]{1999Ap&SS.265..265F} Fuhrmann, K.\ 1999, \apss, 
265, 265 

\bibitem[\protect\citeauthoryear{Fuhrmann}{2004}]{2004AN....325....3F} 
Fuhrmann K., 2004, AN, 325, 3 



\bibitem[\protect\citeauthoryear{Gilli et al.}{2006}]{2006A&A...449..723G} 
Gilli G., Israelian G., Ecuvillon A., Santos N., Mayor M., 2006, A\&A, 
449, 723 

\bibitem[\protect\citeauthoryear{Haywood}{2001}]{2001MNRAS.325.1365H} Haywood M., 2001, MNRAS, 325, 1365 


\bibitem[\protect\citeauthoryear{Haywood}{2006}]{2006MNRAS.371.1760H} Haywood M., 2006, MNRAS, 371, 1760 

\bibitem[\protect\citeauthoryear{Haywood}{2007}]{}Haywood M., 2007, IAU Symposium 248, A giant step: from 
milli to microarcsecond astrometry.

\bibitem[\protect\citeauthoryear{Ida \& Lin}{2005}]{2005PThPS.158...68I} 
Ida S., Lin D.~N.~C., 2005, PThPS, 158, 68 




\bibitem[\protect\citeauthoryear{J{\o}rgensen \& 
Lindegren}{2005}]{2005A&A...436..127J} J{\o}rgensen B.~R., Lindegren L., 
2005, A\&A, 436, 127 


\bibitem[\protect\citeauthoryear{Laughlin}{2000}]{2000ApJ...545.1064L} 
Laughlin G., 2000, ApJ, 545, 1064 



\bibitem[\protect\citeauthoryear{Mordasini et 
al.}{2007}]{2007arXiv0710.5667M} Mordasini C., Alibert Y., Benz W., Naef 
D., 2007, arXiv, 710, arXiv:0710.5667 



\bibitem[\protect\citeauthoryear{Nordstr{\"o}m et al.}{2004}]{2004A&A...418..989N} Nordstr{\"o}m B., et al., 2004, A\&A, 418, 989 


\bibitem[\protect\citeauthoryear{Reddy et al.}{2003}]{2003MNRAS.340..304R} 
Reddy B., Tomkin J., Lambert D., Allende Prieto C., 2003, MNRAS, 340, 
304 


\bibitem[\protect\citeauthoryear{Reddy, Lambert, \& Allende Prieto}{2006}]{2006MNRAS...367.1329} Reddy B., Lambert D., Allende Prieto C., 2006, MNRAS, 367, 1329


\bibitem[\protect\citeauthoryear{Robinson et al.}{2006}]{2006ApJ...643..484R} Robinson S.~E., Laughlin G., Bodenheimer 
P., Fischer D., 2006, ApJ, 643, 484 


\bibitem[Santos et al.(2000)]{2000A&A...363..228S} Santos, N.~C., 
Israelian, G., \& Mayor, M.\ 2000, \aap, 363, 228 



\bibitem[Santos et al.(2007)]{2007A&A...474..647S} Santos, N.~C., Mayor, M., Bouchy, F., et al.\ 2007, \aap, 474, 647 


\bibitem[\protect\citeauthoryear{Sellwood \& 
Binney}{2002}]{2002MNRAS.336..785S} Sellwood J.~A., Binney J.~J., 2002, 
MNRAS, 336, 785 

\bibitem[\protect\citeauthoryear{Udry \& 
Santos}{2007}]{2007ARA&A..45..397U} Udry S., Santos N.~C., 2007, ARA\&A, 
45, 397 



\bibitem[\protect\citeauthoryear{Valenti \& 
Fischer}{2005}]{2005ApJS..159..141V} Valenti J.~A., Fischer D.~A., 2005, 
ApJS, 159, 141 


\end{thebibliography}
\end{document}